\documentclass[review]{elsarticle}

\usepackage{graphicx}
\usepackage{booktabs}
\usepackage{amsmath}
\usepackage{lineno,hyperref}
\usepackage{caption}
\usepackage{subcaption}
\usepackage{setspace}
\usepackage{multicol}
\usepackage{float}
\usepackage[british]{babel}
\usepackage{amsmath,amssymb,bm}

\usepackage[margin=1in]{geometry}

\modulolinenumbers[5]

\begin{document}

%\singlespacing 

\begin{frontmatter}        

%\title{The Zero-adjusted Poisson-Weibull for fraud detection in loan portfolios}
\title{The zero-inflated promotion cure rate regression model applied to fraud propensity in bank loan applications}

%\title{Elsevier \LaTeX\ template\tnoteref{mytitlenote}}


\author{Francisco Louzada\fnref{myfootnote}}
\address{Institute of Mathematical Science and Computing at the University of S{\~a}o Paulo (USP), Brazil} 
\fntext[myfootnote]{Corresponding author: louzada@icmc.usp.br}

\author{Mauro R. de Oliveira Jr.}
\address{Caixa Econ{\^o}mica Federal and Federal University of S{\~a}o Carlos, Brazil}
%\fntext[myfootnote]{Corresponding author: mauroexatas@gmail.com}
%

\author{Fernando F. Moreira}
\address{Credit Research Centre, University of Edinburgh Business School, Scotland, UK}

%
%\author{Mauro R. de Oliveira Jr.\fnref{myfootnote}}
%\address{Caixa Econ{\^o}mica Federal and Federal University of S{\~a}o Carlos, Brazil}
%\fntext[myfootnote]{Corresponding author: mauroexatas@gmail.com}
%
%\author{Francisco Louzada}
%\address{Institute of Mathematical Science and Computing at the University of S{\~a}o Paulo (USP), Brazil} 
%%\fntext[myfootnote]{Corresponding author: louzada@icmc.usp.br}
%
%%
%
%\author{Fernando F. Moreira}
%\address{University of Edinburgh Business School, Scotland, UK}
%%
%%\author{Raffaella Calabrese}
%%\address{University of Essex Business School}




\begin{abstract}
 
In this paper we extend the promotion cure rate model proposed by \citet{Chen99}, by incorporating excess of zeros in the modelling. Despite allowing to relate the covariates to the fraction of cure, the current approach, which is based on a biological interpretation of the causes that trigger the event of interest, does not enable to relate the covariates to the fraction of zeros. The presence of zeros in survival data, unusual in medical studies, can frequently occur in banking loan portfolios, as presented in \citet{louzada2015zero}, where they deal with propensity to fraud in lending loans in a major Brazilian bank. To illustrate the new cure rate survival method, the same real dataset analyzed in \citet{louzada2015zero} is fitted here, and the results are compared.  
\end{abstract}

\begin{keyword}
\texttt {bank loans \sep fraud detection  \sep portfolios \sep survival \sep  zero-inflated}
%\texttt{elsarticle.cls}\sep \LaTeX\sep Elsevier \sep template
%\MSC[2010] 00-01\sep  99-00
\end{keyword}

\end{frontmatter}

%\linenumbers

\section{Introduction}\label{introd}

The cure rate model overcomes the disadvantage of the standard model for survival analysis, where all individuals are susceptible to the occurrence of the event of interest. To handle this problem, \citet{berkson1952survival} proposed a simple model that added the fraction of cured $(p>0)$ into the survival analysis, getting the following expressions for the survival, and density, functions:

\begin{eqnarray}\label{crm}
S(t)   & =  & p + (1-p)S^{*}(t), \ \ \ \ \ t \geq 0,      \\
f(t)   & =  & (1-p)f^{*}(t), \ \ \ \ \ \ \ \ \ \ t \geq 0,
\end{eqnarray}

where $S^{*}$ is the survival function of the subject susceptible to failure, $f^{*}$ is the density probability function, and $p$ is the proportion of subjects immune to failure (cured). This model is called cure rate model, or long-term survival model.

Therefore, $S$ is an improper survival function,  unlike $S^{*}$, since it satisfies:

\begin{equation}
	\displaystyle \lim_{t \to \infty}S(t)=p > 0.
\end{equation}

The advantage of the cure rate model is that it allows to associate covariates in both parts of the model, i.e., it allows covariates to have different influence on cured patients, linking them with $p$, and on patients who are not cured, linking them with parameters of the proper survival function $S^{*}$. 

To accommodate the presence of zeros, which is impossible in the cure rate model shown above, \citet{louzada2015zero} proposed a zero-inflated cure rate model, which survival function is given by:

\begin{equation}
	S(t)= \gamma_1 + (1-\gamma_0-\gamma_1)S^{*}(t), \ \ \ \ \ \ t \geq 0,
	\label{zicrm}
\end{equation}

where $S^{*}$ is the survival function related to the $(1-\gamma_0-\gamma_1)$ proportion of subject susceptible to failure, $\gamma_0$ is the proportion of zero-inflated survival times, and $\gamma_1$ is the proportion of subjects immune to failure (cured or long-term survivors).

Thus, it is now possible to link together the influence of the covariates in the three parts of the model, i.e., to the proportion of the zero-inflated survival times, whose authors identified as fraudulent clients, along with the usual sub-populations of susceptible and non-susceptible to the event of interest. In that paper, the event of interest was related to the time span until the occurrence of default on bank loan portfolios.

The fact that differentiates the zero-inflated cure version from the standard cure approach is expressed in the second of the following satisfied properties: 

\begin{equation}
	\displaystyle \lim_{t \to \infty}S(t)=\gamma_1 > 0.
\end{equation}

\begin{equation}
	\displaystyle S(0)=1-\gamma_0 < 1.
\end{equation}

Note that, if $\gamma_0=0$, i.e., without the excess of zeros, we have the cure rate model of \citet{berkson1952survival}.

\begin{figure}[htbp]
	\centering
		\includegraphics[scale=0.15]{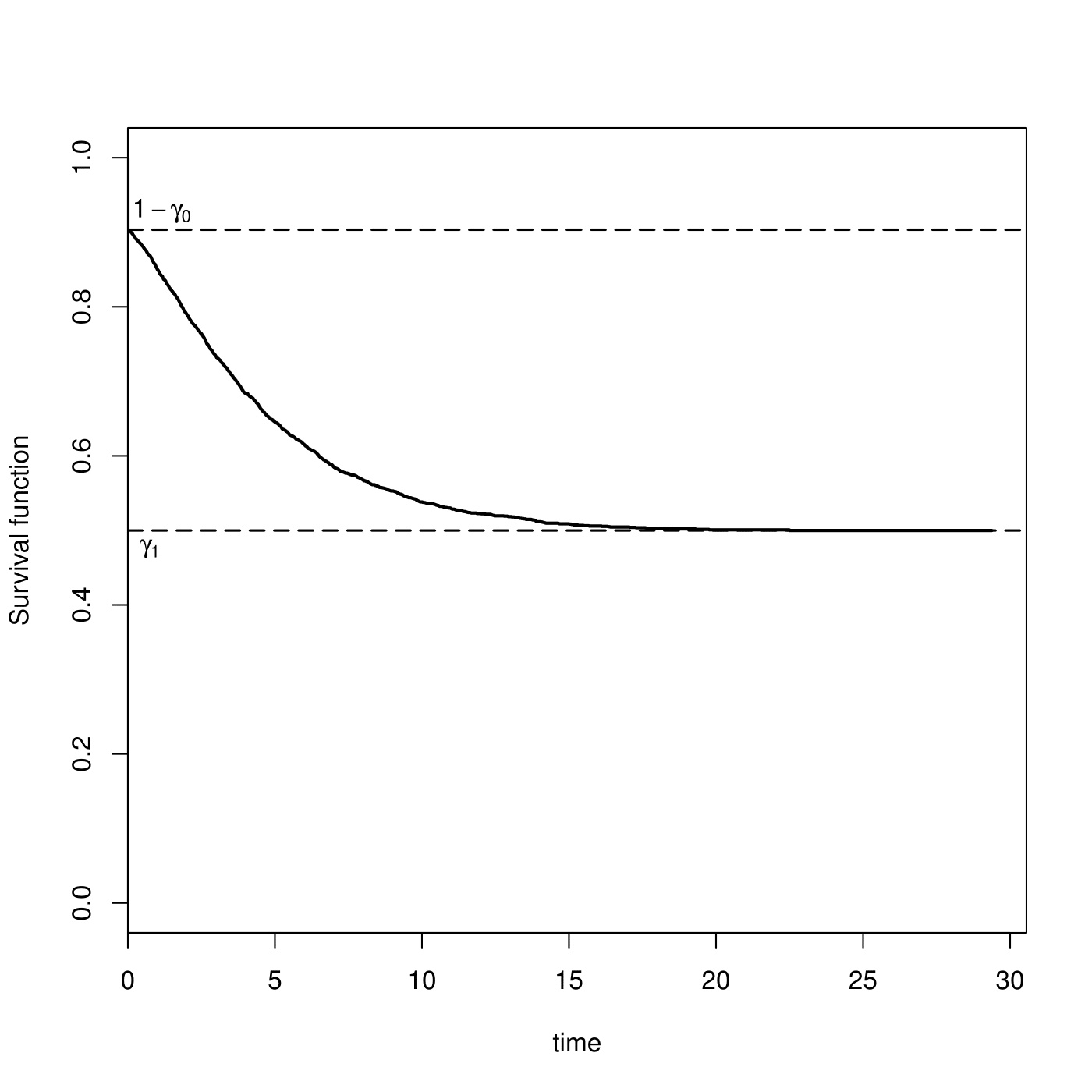}
	\caption{Survival function of the zero-inflated cure rate model as presented in \citet{louzada2015zero}.}
	\label{fig:fig1_LD_ZAW}
\end{figure}

\subsection{Preliminary}                    

In this section we shall briefly describe the promotion cure rate model presented by \citet{Chen99}, further extended by \citet{rodrigues2009unification}, wherein we follow the same notations. This model also incorporates the presence of immune individuals to the event of interest, but still has the disadvantage of not accommodating zero time in its framework.

This survival model with fraction of cure, according to \citep{Chen99}, is based on a biological interpretation of the causes that trigger (promote) a cancer disease relapse. As described by the authors, the process that leads to formation of a detectable cancer mass is triggered by a set of $N$ competitive underlying causes, biologically represented by the number of carcinogenic cells that the individual has left active after the initial treatment. In their paper, it is assumed that $N$ follows a Poisson distribution with mean $\theta$. 

Regarding to the time until the relapse of the cancer under treatment, the authors in \citep{Chen99} have let $Z_i$ be the random time for the $ith$ carcinogenic cells to produce a detectable cancer mass, i.e., the incubation time for the $ith$ (out of $N$) carcinogenic cell. The random variables $Z_i$, $i  =  1, 2,\cdots$, are assumed to be iid, with a common distribution function $F(t)  =  1 -  S(t)$, and are independent of $N$. 

In order to include those individuals who are not susceptible to the event of cancer relapse, i.e, the individuals with the initial number of cancer cells, $N$, equals to $0$ and, theoretically, with infinity survival time, it is assumed that $P(Z_0 =  \infty) =  1$.

Finally, the time to the relapse of cancer is defined by the random variable $T =  min\{Z_i, 0\leq i \leq  N\}$, and therefore, the survival function of $T$, for the entire population, is given by 

\begin{eqnarray}\label{bcrm}
S_p(t) & =  & P(T>t, N \geq 0)     \nonumber \\
       & =  & P(N=0) + P(Z_1>t, \cdots, Z_N>t, N \geq 1)   \nonumber \\
       & =  & \exp(-\theta) + \sum_{k=1}^{\infty} {S(t)}^k \frac{{\theta}^k}{k!}\exp(-\theta) \nonumber \\
		   & =  & \exp(-\theta +\theta S(t)) = \exp(-\theta F(t)). 		
\end{eqnarray}

The density function corresponding to \ref{bcrm} is given by 

\begin{eqnarray}\label{dfbcrm}
f_p(t)  & =  &  \theta f(t) \exp(-\theta F(t)), 		
\end{eqnarray}

We notice that, $S_p$ and $f_p$ are not, properly, survival function and density function, respectively. In fact, note that, $P(Z_0 =  \infty) =  1$, leads to the proportion $\displaystyle{\lim_{t \to \infty}S_p(t) \equiv S_p(\infty)  \equiv P(N=0) = \exp(-\theta) >0}$, from the population of individuals who are not susceptible to the occurrence of cancer relapse (cured). Moreover, the fraction of cure is very flexible, i.e., it has the property to accommodate a wide variety of cases, since as $\theta \rightarrow \infty$, the proportion of cured tends to $0$, whereas as $\theta \rightarrow 0$, the proportion of cured tends to $1$.

In the situation where we consider the model formulation taking into account only susceptible individuals, that is, when it is present in all individuals a number of initial cancer cells greater than zero, we have a slightly modified expression for the survival function:

\begin{eqnarray}\label{nonbcrm}
S^{*}_p(t) & =  &  P(T>t, N \geq 1)   = \frac{\exp(-\theta F(t))-\exp(-\theta)}{1-\exp(-\theta)}.  
\end{eqnarray}

According to the formulation, we found now that $S^{*}_p(t)$ is a proper survival function, since the following conditions are satisfied: $S^{*}_p(0)=1$ and $S^{*}_p(\infty)=0$. Still following the model presentation as in \citep{Chen99}, we come to the probability density function of individuals who are susceptible to recurrence of the considered event: 

\begin{eqnarray}\label{dennonbcrm} 
f^{*}_p(t) & =  &  -\frac{d}{dt}S^{*}(t)   = \left( \frac{\exp(-\theta F(t))}{1-\exp(-\theta)} \right) \theta f(t). 
\end{eqnarray}

Finally, we come to the mathematical relation between the cure rate model, as presented by \citet{berkson1952survival}, equation \ref{crm}, and the biological based model, as proposed by \citet{Chen99}, equation \ref{bcrm},

\begin{eqnarray}\label{relati}
S_p(t)  & =  & \exp(-\theta) + (1-\exp(-\theta))S^{*}_p(t), \ \ \ \ \ \ t \geq 0,		\\ 
f_p(t)  & =  & (1-\exp(-\theta))f^{*}_p(t), \ \ \ \ \ \ \ \ \ \ \ \ \ \ \ \ \ \ \ \ \ t \geq 0,	
\end{eqnarray}

where $S^{*}_p$ and $f^{*}_p$, are given by the proper survival function, and proper density function, as in \ref{nonbcrm} and \ref{dennonbcrm}, respectively. Thus, we see that the {\citet{Chen99}}`s model can be rewritten as a cure rate model, with cure rate equal to $p = \exp(-\theta)$.

\subsection{Motivation}

Although the promotion model, given by \ref{bcrm}, is formulated within a biological context, it has also been applied in other areas, such as credit risk analysis of bank loan portfolios. In these new developments, the number $N$ is related to the number of risks that compete to the occurrence of a particular financial event of interest, i.e., default or non-performing of loans. Therefore, the formulation admits generalizations in various ways, such as done in \citet{barriga2015non}, where the authors studied the time until the event of default on a Brazilian personal loan portfolio, and where the authors let $N$ follow a geometric distribution, and $F(t)$ be a cumulative density function of the inverse Weibull distribution.

Also in the area of credit risk modelling in \citet{oliveira2014serasaeng}, the authors applied the model given by \ref{bcrm} to study the time span to full recovery of non-performing loans in a portfolio of personal loans. In \citet{oliveira2014evidence} the authors compare the parameters $\theta$ obtained from two follow-up studies of a set of non-performing loans. The first follow-up is related to the time span to the default occurrence, while the second one is related to the time until the full recovery of the related defaulted loans. The authors found a significant relationship between default and recovery processes. The article suggests that in times of higher risk of default, it is also likely to have a decrease in the recovery rates of non-performing loans.

\subsection{Proposal}

To accommodate zero excess in a survival analysis of loan portfolios, in \citep{louzada2015zero}, the authors have proposed a modification in the survival function of the cure rate model, which has led to the improper survival function given in \ref{zicrm}, also labelled as zero-inflated cure rate model. In this scenario, information from fraudsters in loan applications is exploited through the joint modelling of the survival time of fraudsters, which is equal to zero, along with the survival times of the remaining portfolio.

The purpose of this article is to follow the method as done in \citep{louzada2015zero}, and then propose a way of incorporating the fraction of zeros into the biological based model of \citet{Chen99}. The advantage of incorporating the fraction of zeros in this model is that the model enables us to interpret the underlying causes that lead to occurrence of default so that we can compare if customers at higher risk of default are also more likely to commit fraud in credit applications.
 
Such an approach leads the credit risk management to a complete overview of the risk factors involved in lending, that is, dealing with fraud prevention, non-performing loan control and ensure customer loyalty. To exemplify the application of the proposed approach, we analyse a portfolio of loans made available by a large Brazilian commercial bank. 

The paper is organized as follows. In Section \ref{model}, we formulate the new model named zero-inflated promotion cure rate, and we present the approach for parameter estimation. An application to a real data set is presented in Section \ref{apl}. Some general remarks are presented in Section \ref{con}.

\section{Model specification}\label{model}

In what follows, we consider the promotion cure rate model as defined in equation \ref{relati}. Hence, we propose a new (improper) survival function as following:

\begin{equation}
	S_p(t)= \gamma_1 + (1-\gamma_0-\gamma_1)S^{*}_p(t), \ \ \ \ \ \ t \geq 0,
	\label{newzicrm}
\end{equation}

where $S^{*}_p$ is given by \ref{nonbcrm}, and the parameters $\gamma_0$ and $\gamma_1$ are defined as following: $$\gamma_0 = \exp(-\kappa)$$ and $$\gamma_1 = \exp(-\theta).$$

To ensure that $\gamma_0$, $\gamma_1$, and $(1-\gamma_0-\gamma_1) \in (0,1)$, we  propose to link two q-dimensional vector of covariates, $x_1$ and $x_2$, into the parameters related to zero inflation and cure, respectively, as following:

	\begin{equation}
 \left\{
\begin{array}{llll}
\displaystyle  \kappa_{i} &=& - \log \left(\frac{e^{\mathbf x^\top_{1i}\beta_{\kappa}}}{1+e^{\mathbf x^\top_{1i}\beta_{\kappa}}+e^{\mathbf x^\top_{2i}\beta_{\theta}}}\right),\\

\\

\displaystyle  \theta_{i}  &=& - \log \left(\frac{e^{\mathbf x^\top_{2i}\beta_{\theta}}}{1+e^{\mathbf x^\top_{1i}\beta_{\kappa}}+e^{\mathbf x^\top_{2i}\beta_{\theta}}}\right),\\

\end{array}
\right.
\label{linkr}
\end{equation}

where $\beta_{\kappa}\in \Re^q$, is a q-dimensional vector of regression coefficients to be estimated, that relates the influence of the covariates into the excess of zeros, while  $\beta_{\theta}\in \Re^q$, is a q-dimensional  vector of regression coefficients that relates the influence of the covariates into the fraction of cured.

To complete the configuration of the model, i.e., to determine the parametric form of $S^{*}_p$, we let $f(t)$ and $F(t)$ be, respectively, the density probability function and the cumulative probability function of the Weibull distribution. The Weibull distribution is a continuous probability distribution, commonly applied in survival analysis and reliability. It has two parameters, $\alpha > 0$ and $\lambda > 0$, respectively, the shape and scale parameters. 

Finally, we come to the following model framework:

	\begin{equation}
 \left\{
\begin{array}{llll}
	S_p(t) & = & \exp(-\theta) + (1-\exp(-\kappa)-\exp(-\theta))S^{*}_p(t), \\
	S^{*}_p(t) & =  & \frac{\exp(-\theta F(t))-\exp(-\theta)}{1-\exp(-\theta)}, \\  
  f^{*}_p(t) & =  & \left( \frac{\exp(-\theta F(t))}{1-\exp(-\theta)} \right) \theta f(t), \\ 
  F(t)       & =  & 1 - e^{- \left( \frac{t}{\lambda} \right)^\alpha}, \\
	f(t)       & =  & \frac{\alpha}{\lambda} {\left( \frac{t}{\lambda} \right)}^{\alpha-1} e^{\left( - \frac{t}{\lambda} \right)^\alpha}.
\end{array}
\right.
\label{modlinkr}
\end{equation}

%where $\Gamma(\cdot)$ denotes the gamma function, and $\gamma(\cdot)$ denotes the lower incomplete gamma function and, finally,  $a > 0$, $b > 0$, and $k > 0$, refer to the three parameters of the generalized gamma distribution.

\subsection{Likelihood function}

Regarding to the likelihood contribution of each consumer survival time $t_i$, we note that there are different sub-group of clients. Therefore, the likelihood contribution of a client $i$, linked with a par of q-dimensional vector of covariates, $x_1$ and $x_2$, as in \ref{linkr}, hence, our proposed zero-inflated survival model must assume three different values:

\begin{equation}
\left\{
\begin{array}{lcl}
\gamma_{0i},& \mbox{if} &  t_i= 0,\\
(1 - \gamma_{0i} - \gamma_{1i})f^{*}_p(t_i), & \mbox{if} & t_i \mbox{ is non right censored}  \\
\gamma_{1i} + (1-\gamma_{0i}-\gamma_{1i})S^{*}_p(t_i), & \mbox{if} & t_i \mbox{ is right censored}.
\end{array}
\right.
\label{ziwpdf}
\end{equation}

Let the data take the form $\mathcal{D} = \left\{{t_{i},\delta_{i}},x_{1i},x_{2i}\right\}$, where $\delta_i=1$ if $t_i$ is an observable time to default, $\delta_i=0$ if it is right censored, for $i = 1,2,\cdots n,$ and $x_1$ and $x_2$ are a couple of vector of covariates associated with a consumer $i$. As we shall see in application section, the variable vectors can be the same, i.e., $x_1=x_2$. Let $(\alpha,\lambda)$ denote the parameter vector of the Weibull distribution and, finally, let $(\beta_{\kappa},\beta_{\theta})$ be the regression parameters associated, respectively, with the proportion of inflation of zeros and the proportion of long-term survivors (cure rate). 

The likelihood function of the proposed new zero-adjusted long-term survival model, with a parameter vector, $\vartheta = (\alpha,\lambda,\beta_{\kappa},\beta_{\theta})$, to be estimated via MLE approach, is based on a sample of $n$ observations, $\mathcal{D} = \left\{{t_{i},\delta_{i},,x_{1i},x_{2i}}\right\}$. Consider the indicator function $I(T_i>0)$, where $I(T_i>0)=1$, if $T_i>0$, and $I(T_i>0)=0$, otherwise. Then, we can write the likelihood function, under non-informative censoring, as 

\begin{equation}
L(\vartheta ; \mathcal{D}) \propto  \prod_{i=1}^n \left\{ (1-I(t_i>0)) \gamma_{i0} + I(t_i>0) {\left[(1 - \gamma_{i0} - \gamma_{i1})f_w^* (t_i)\right]}^{\delta_i} {\left[\gamma_{i1} + (1-\gamma_{i0}-\gamma_{i1})S^{*}(t_i)\right]}^{1-\delta_i}  \right\} 
\label{like}
\end{equation}

\subsection{Parameter estimation}\label{paramest}

Parameter estimation is performed by straightforward use of maximum likelihood estimation (MLE), where, as we will see, its simple application is supported by our simulation studies.

The maximum likelihood estimates $\hat{\vartheta}$, regarding the parameter vector $\vartheta$, both considering the model with or without covariates, are obtained through maximization of $L(\vartheta;\mathcal{D})$ or $\ell(\vartheta;\mathcal{D})=\log\{L(\vartheta;\mathcal{D})\}$. Confidence intervals for the parameters is based on asymptotic normality. According to \citep{migon2014statistical}, under suitable regularity conditions, the asymptotic distribution of the maximum likelihood estimates (MLEs), $\hat{\vartheta}$, is a multivariate normal with mean vector $\vartheta$ and covariance matrix $\textbf{I}^{-1}(\hat{\vartheta})$ , which can be estimated by the observed information matrix $\textbf{I}(\vartheta) = \{-\partial^2\ell(\vartheta)/\partial{\vartheta}\partial{\vartheta}^T\}^{-1}$, evaluated at $\vartheta=\hat{\vartheta}$. Therefore, let $I^{ii}$ be the $ith$ diagonal element of the inverse of $\textbf{I}$, evaluated in $\hat{\vartheta}$. An approximate $1000(1-\alpha)\%$ confidence interval for $\hat{\vartheta_i}$, based on assumed regularity conditions, is given by $\left(\hat{\vartheta_i} - z_{\alpha/2}\sqrt{I^{ii}}, \ \ \hat{\vartheta_i} + z_{\alpha/2}\sqrt{I^{ii}} \right)$,  where $z_{\alpha}$ denotes the 
$1000(1-\alpha)$ percentile of the standard normal random variable. In the application section we set $\alpha=0.05$, where we get a $95\%$ confidence interval for each ML estimate.

There are various software and routines available to approximate the parameter estimate and confidence interval described above. We choose the method ``BFGS", see details in \citep{manualR}, which comes within the \textbf{R} routine \texttt{optim}.

\section{Bank loan survival data}\label{apl}

In this section we present an application of the zero-inflated promotion cure rate model in a database made available by one of the largest Brazilian banks. The data comprises bank lending survival times, i.e., the time span between the grant and the moment that the customer fails to honour its payment back to a bank. 

Therefore, we have a set of non-negative times to model. Our goal is to confirm whether a characteristic of a group of customers makes it more likely to fraud in bank loan applications. A client is classified as a fraudster if, after the loan granting, she or he does not pay any commitment thereof, so her/his survival time is zero compared to other borrowers in the portfolio.

It is important to note that the given database, quantities, rates, and levels of the available covariate, do not necessarily represent the actual condition of the financial institution's customer base. Despite being a real database, the bank may have sampled the data in order to change the current status of its loan portfolio.

Figures \ref{sumbase1} presents a summary of the portfolio together with the portfolio subdivision considering one available covariate. For reasons of confidentiality, we will refer to it as covariate $x$. In this case, we can see that $x$ has three levels, referring to a particular characteristics of a bank's customer profile. Note that, it is slightly different than one presented in \citep{louzada2015zero}, p. 14 , because it is a sample extracted randomly from the database provided by the bank.

\begin{table}[H]
\centering
\caption{Summary of the bank loan lifetime data.}
%\smallskip\noindent
%\resizebox{\linewidth}{!}{%
\begin{tabular}{|ccccc|}
\toprule
Portfolio 		 & Number of 		 & Number of      &  Number of 			& Number of	   \\ 
							 & consumers   	 & fraudsters     &  defaulters  		& censored	   \\ 
\hline
$x=1$          & 1,626  			 & 137 (8.4256\%)	&	305 (18.7577\%)	& 1,184 (72.8167\%)\\
$x=2$          & 1,574  			 & 127 (8.0686\%)	&	242 (15.3748\%)	& 1,205 (76.5565\%)\\
$x=3$          & 938   				 &  30 (3.1983\%)	&	 93 (9.9147\%)	&   815 (86.8870\%)\\
\hline
Total					 & 4,138 				 & 294 (7.1049\%)	&	640 (15.4664\%)	& 3,204 (77.4287\%)\\

\bottomrule
   \end{tabular}\label{sumbase1}
%}
\end{table}

The following are two applications, the first related to the proposed model by \ref{modlinkr}, and the second, for comparison purposes, referring to the proposed model by \citet{louzada2015zero}.

\subsection{Application 1: Zero-inflated promotion cure rate model}\label{model1}
                                           
To proceed with the applications, we will deal with dummy variables. As $x$ has three levels, then we have two dummy variables, $dx1$ and $dx2$, where $dx1=1$, if $x=1$, and $dx1=0$, otherwise, and similarly, $dx2=1$, if $x=2$, and $dx2=0$, otherwise. The customer group such that $x = 3$ is characterized by setting  $dx1$~$=$~$dx2$~$=0$.
				
According to \ref{aplimod1}, we do not link covariates to the $(\alpha,\lambda)-$  Weibull parameters, which we leave for future research. However, note that we made the following re-parametrization, $\alpha=e^{\alpha^{'}}$ and  $\lambda=e^{\lambda^{'}}$. Thus we have the following set of parameters $\left\{\beta_{10}, \beta_{11}, \beta_{12}, \beta_{20}, \beta_{21}, \beta_{22},\alpha^{'}, \lambda^{'}\right\}$ to be estimated by MLE, as specified in section \ref{paramest}.

\begin{equation}
 \left\{
\begin{array}{cccc}
\displaystyle  \kappa_{i} &=& -\log\left(\frac{e^{\mathbf \beta_{\kappa0} + {dx1}_i\beta_{\kappa1}+ {dx2}_i\beta_{\kappa2}}}{1+e^{\mathbf \beta_{\kappa0} + {dx1}_i\beta_{\kappa1}+ {dx2}_i\beta_{\kappa2}}+e^{\mathbf \beta_{\theta0} + {dx1}_i\beta_{\theta1}+ {dx2}_i\beta_{\theta2}}}\right),\\

\displaystyle  \theta_{i}  &=& -\log\left(\frac{e^{\mathbf \beta_{\theta0} + {dx1}_i\beta_{\theta1}+ {dx2}_i\beta_{\theta2}}}{1+e^{\mathbf \beta_{\kappa0} + {dx1}_i\beta_{\kappa1}+ {dx2}_i\beta_{\kappa2}}+e^{\mathbf \beta_{\theta0} + {dx1}_i\beta_{\theta1}+ {dx2}_i\beta_{\theta2}}}\right),\\

\alpha               &=& e^{\alpha^{'}}, \\

\lambda               &=& e^{\lambda^{'}}. \\

\end{array}
\right.
\label{aplimod1}
\end{equation}

\begin{table}[H]
\centering
\caption{Maximum likelihood estimation results for the zero-inflated Promotion cure rate regression model.}
\begin{tabular}{|cccc|}
\toprule
Parameter& Estimate (est) & Standard error (se) & $|$est$|/$ se  \\ 
\hline
$	\beta_{\kappa0}	$		&	-1,4108	&	0,2132	&	6,6165	\\
$	\beta_{\kappa1}	$		&	0,3832	&	0,2333	&	1,6424	\\
$	\beta_{\kappa2}	$		&	0,5245	&	0,2363	&	2,2195	\\
$	\beta_{\theta0}	$		&	1,8575	&	0,1208	&	15,3816	\\
$	\beta_{\theta1}	$		&	-0,8011	&	0,1296	&	6,1823	\\
$	\beta_{\theta2}	$		&	-0,5504	&	0,1328	&	4,1460	\\
$	\alpha^{'}	$				&	0,1337	&	0,0438	&	3,0557	\\
$	\lambda^{'}	$				&	3,2746	&	0,0872	&	37,5577	\\
\bottomrule
   \end{tabular}\label{LT7}
\end{table}

From Table \ref{LT7}, through the analysis of the parameter significance, we see that the available variable $x$ is very significant to discriminate risk of fraud, default risk, and the cure rate.

%we see that being part of group $x=3$ is significant for differentiation on the propensity to fraud, where customers with this characteristic are less likely to be fraudster as customers of other groups.

The parameters $\beta_{\kappa0}$, $\beta_{\kappa1}$	and $\beta_{\kappa2}$, confirm that the segmentation given by the covariable $x$ is significant to establish an ascending order of fraud rates, from the group $x=3$, with the lowest fraud rate group, to the highest fraud rate group, $x=1$.

On the other hand, the parameters $\beta_{\theta0}$, $\beta_{\theta1}$ and $\beta_{\theta2}$, confirm that the segmentation given by the covariate $x$ is significant to establish a descending order of long-term survival rates, from the group $x=3$, with higher cure rate, to the lowest cure rate group, $x=1$.

\subsection{Application 2: Zero-inflated cure rate model}\label{model2}

Similarly, according to \ref{aplimod2}, we do not link covariates to the $(\alpha,\lambda)$ Weibull parameters and, again, note that we made the same re-parametrization, $\alpha=e^{\alpha^{'}}$ and  $\lambda=e^{\lambda^{'}}$. Thus, we have now the following set of parameters $\left\{\beta_{10}, \beta_{11}, \beta_{12}, \beta_{20}, \beta_{21}, \beta_{22},\alpha^{'}, \lambda^{'}\right\}$, to be estimated by MLE.

\begin{equation}
 \left\{
\begin{array}{cccc}
\gamma_{0i} &=& \left(\frac{e^{\mathbf \beta_{10} + {dx1}_i\beta_{11}+ {dx2}_i\beta_{12}}}{1+e^{\mathbf \beta_{10} + {dx1}_i\beta_{11}+ {dx2}_i\beta_{12}}+e^{\mathbf \beta_{20} + {dx1}_i\beta_{21}+ {dx2}_i\beta_{22}}}\right),\\

\gamma_{1i} &=& \left(\frac{e^{\mathbf \beta_{20} + {dx1}_i\beta_{21}+ {dx2}_i\beta_{22}}}{1+e^{\mathbf \beta_{10} + {dx1}_i\beta_{11}+ {dx2}_i\beta_{12}}+e^{\mathbf \beta_{20} + {dx1}_i\beta_{21}+ {dx2}_i\beta_{22}}}\right),\\

\alpha               &=& e^{\alpha^{'}}, \\

\lambda               &=& e^{\lambda^{'}}, \\

\end{array}
\right.
\label{aplimod2}
\end{equation}

\begin{table}[H]
\centering
\caption{Maximum likelihood estimation results for the zero-inflated Weibull cure rate regression model.}
\begin{tabular}{|cccc|}
\toprule
Parameter& Estimate (est) & Standard error (se) & $|$est$|/$ se  \\ 
\hline
$\beta_{10}$	& 	-1,3894033	& 	0,21264846	& 	6,533803725	\\
$\beta_{11}$	& 	0,3657269	  & 	0,23363029	& 	1,565408749	\\
$\beta_{12}$	& 	0,5130939	  & 	0,23665962	& 	2,168066948	\\
$\beta_{20}$	& 	1,8784957	  & 	0,11970854	& 	15,69224468	\\
$\beta_{21}$	& 	-0,8121232	& 	0,13204012	& 	6,150579082	\\
$\beta_{22}$	& 	-0,5585139	& 	0,13502802	& 	4,136281492	\\
$\alpha^{'}$	& 	0,1121479	  & 	0,04331601	& 	2,589063489	\\
$\lambda^{'}$ & 	3,1692889	  & 	0,07672007	& 	41,30977592	\\
\bottomrule
   \end{tabular}\label{LT8}
\end{table}

The analysis of the parameter significance in the \ref{LT8} shows that the available variable $x$ is very significant to discriminate risk of fraud, default risk, and the cure rate.

The parameters $\beta_{10}$, $\beta_{11}$	and $\beta_{12}$, reconfirm that the segmentation given by the covariable $x$ is significant to establish an ascending order of fraud rates, from the group $x=3$, with the lowest fraud rate group, to the highest fraud rate group, $x=1$.

On the other hand, the parameters $\beta_{20}$, $\beta_{21}$	and $\beta_{22}$, reconfirm that the segmentation given by the covariable $x$ is significant to establish a descending order of long-term survival rates, from the group $x=3$, with higher cure rate, to the lowest cure rate group, $x=1$.

\subsection{Some results and comparisons.}\label{resultscompa}

\begin{figure}[H]
	\centering
		\includegraphics[scale=0.2]{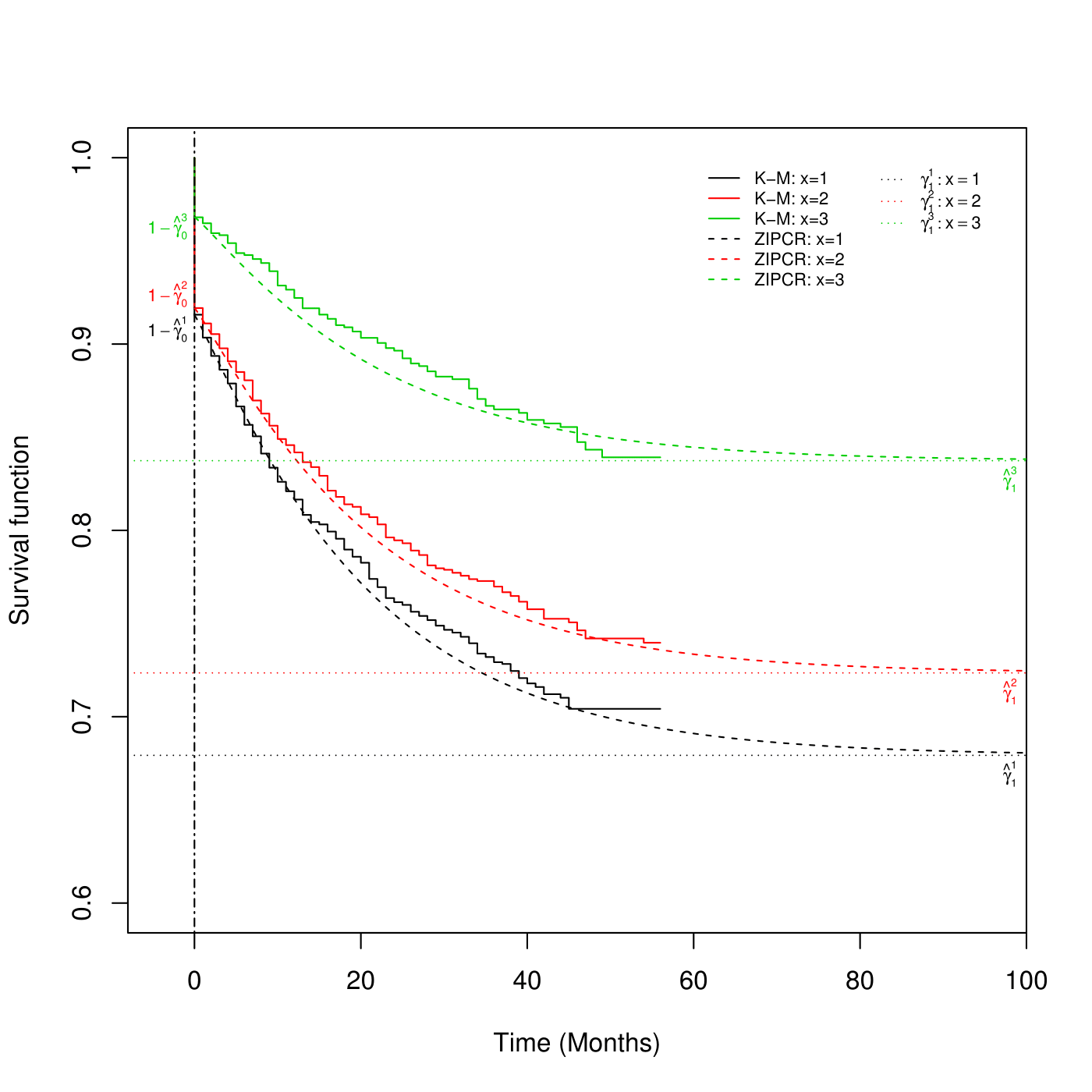}
	\caption{The Kaplan-Meier (K-M) survival curve, according to customer profile $x=1$, $x=2$ and $x=3$, and survival functions estimated by the zero-inflated promotion cure rate regression model (ZIPCR).}
	\label{ajuste_LD_ZAW}
\end{figure}

The Kaplan-Meier (K-M) survival curves for the Brazilian bank loan portfolio used in this study, according to customer profiles $x=1$, $x=2$ and $x=3$, are presented in the Figure \ref{ajuste_LD_ZAW}. This figure presents the fitted survival functions, as defined by the zero-inflated promotion cure rate (ZIPCR) regression model in \ref{modlinkr}. We observed that the improper survival function falls instantly, at $t = 0$, to the values ${1-\hat{\gamma}_0}^1$, ${1-\hat{\gamma}_0}^2$, ${1-\hat{\gamma}_0}^3$ (see results in the Table \ref{results1}), respectively, when $x = 1$, $x = 2$ or $x = 3$. So, the model accommodated well the zero-inflated survival times and, as expected, the plateaus at points ${\hat{\gamma}_1}^1$, ${\hat{\gamma}_1}^2$ and ${\hat{\gamma}_1}^3$, highlighted the presence of cure rate.

It is important to note that both parameters, $\beta_{\kappa1}$	 and $\beta_{11}$, were not statistically significant, what can be confirmed in the chart of Kaplan-Meier survival curves stratified by $x$.

Tables \ref{results1} and \ref{results2} show the compared model results. We can see that the outcomes are very satisfactory, when they are compared with the actual data made available by the bank. We can see that the estimated parameters ratify the behaviour of groups, as showed in Table \ref{sumbase1}.

\begin{table}[H]
\centering
\caption{Modelling outcomes of application 1}
%\smallskip\noindent
%\resizebox{\linewidth}{!}{%
\begin{tabular}{|cccc|}
\toprule
Estimated 		 & Consumer         & Consumer    & Consumer     \\ 
parameter			 & with $x=1$       & with $x=2$  & with $x=3$   \\ 
\hline
$\hat{\gamma}_0= \exp(-\hat{\kappa})$       &  8.4526\%       &  8.0701\%    &  3.1884\%    \\
$\hat{\gamma}_1= \exp(-\hat{\theta})$       & 67.9279\%       & 72.3510\%    & 83.7422\%    \\
\bottomrule
   \end{tabular}\label{results1}
%}
\end{table}

\begin{table}[H]
\centering
\caption{Modelling outcomes of application 2}
%\smallskip\noindent
%\resizebox{\linewidth}{!}{%
\begin{tabular}{|cccc|}
\toprule
Estimated 		 & Consumer         & Consumer    & Consumer     \\ 
parameter			 & with $x=1$       & with $x=2$  & with $x=3$   \\ 
\hline
$\hat{\gamma}_0 $      &  8.4255\%       &  8.0687\%    &  3.1981\%    \\
$\hat{\gamma}_1$       & 68.1229\%       & 72.5502\%    & 83.9697\%    \\
\bottomrule
   \end{tabular}\label{results2}
%}
\end{table}

Table \ref{statistic} shows the results for AIC and BIC criteria for both models. 

\begin{table}[H]
\centering
\caption{ Statistics from the adjusted models.}
%\smallskip\noindent
%\resizebox{\linewidth}{!}{%
\begin{tabular}{|lccc|}
\toprule

Model           &     \multicolumn{3}{c|}{Statistic}\\
\cline{2-4}
	
%Model     		                   &                & Statistic    &       \\ 

       			                     & l$(\cdot)$     & AIC          &   BIC  \\ 
\hline
Zero-inflated promotion cure rate     &  -5,035.84       &  10,087.67    &  10,138.3    \\
Zero-inflated cure rate               &  -5,037.44       &  10,090.88    &  10.141.5    \\
\bottomrule
   \end{tabular}\label{statistic}
%}
\end{table}

\section{Concluding remarks}\label{con} 

We introduced a methodology based on a zero-inflated survival data that extends the model proposed by \citet{Chen99}. In this sense, an advantage of our approach is to accommodate zero-inflated times, which is not possible in the standard cure rate model.

To illustrate the methodology presented here, we analyzed a bank loan survival data, in order to assess the propensity to fraud in loan applications. In this scenario, information from fraudsters is exploited through the joint modelling of the survival time of fraudsters, which is equal to zero, along with the survival times of the remaining portfolio.  The results showed the new model performed very well.

Despite the new ZIPCR model presenting slightly better results in terms of AIC and BIC, it is important to note that the actual performance of the new model will be measured through its daily use by the bank and with the use of a wide variety of covariates available, since the model allows the use of as many variables as needed, whether quantitative continuous or categorical.

\section*{Acknowledgement}

\noindent The research was sponsored by CAPES - Process number: BEX 10583/14-9, Brazil.

\section*{References}

\bibliography{Reference}

\end{document}